\documentclass[12pt,preprint]{aastex}

\slugcomment{Accepted for publication in the Astrophysical Journal 
Letters}

\shorttitle{Mid-infrared spectroscopy of submillimeter galaxies}
\shortauthors{Lutz et al.}

\begin{document}

\title{Mid-infrared spectroscopy of two luminous submillimeter 
galaxies at z$\sim$2.8}


\author{D. Lutz\altaffilmark{1}, E. Valiante\altaffilmark{1}, 
E. Sturm\altaffilmark{1}, R. Genzel\altaffilmark{1}, 
L.J. Tacconi\altaffilmark{1}, M.D. Lehnert\altaffilmark{1}, 
A. Sternberg\altaffilmark{2}, A.J. Baker\altaffilmark{3,4}}
\altaffiltext{1}{Max-Planck-Institut f\"ur extraterrestrische Physik, 
Postfach 1312, 85741 Garching, Germany \email{lutz@mpe.mpg.de, 
valiante@mpe.mpg.de, sturm@mpe.mpg.de, genzel@mpe.mpg.de, 
linda@mpe.mpg.de, mlehnert@mpe.mpg.de} }
\altaffiltext{2}{School of Physics and Astronomy, Tel Aviv University, 
Ramat Aviv 69978, Israel \email{amiel@wise.tau.ac.il}}
\altaffiltext{3}{Jansky Fellow, National Radio Astronomy Observatory}
\altaffiltext{4}{University of Maryland, Department of Astronomy, College 
Park, MD 20742-2421 \email{ajb@astro.umd.edu}}

\begin{abstract}
Using the Infrared Spectrograph (IRS) on board the Spitzer Space Telescope, we 
have obtained rest frame mid-infrared spectroscopy of two
bright submillimeter galaxies. SMMJ02399-0136 at z=2.81 
shows a superposition
of PAH emission features and a mid-infrared continuum, indicating 
significant and roughly equal 
contributions to its bolometric luminosity from star formation and from a 
Compton-thick AGN. We derive a 
new redshift of z=2.80 for MMJ154127+6616 from the IRS spectrum and find 
this object is dominated by starburst PAH emission. The rest frame mid- to 
far-infrared spectral energy distributions are consistent with
these submillimeter galaxies being scaled up versions of local 
ultraluminous infrared galaxies.
The mid-infrared spectra support the scenario that submillimeter galaxies 
are sites
of extreme star formation and represent a key phase in the formation of 
massive galaxies.
\end{abstract}


\section{Introduction}

Deep submm and mm surveys using SCUBA and MAMBO have changed our view of 
the early universe by resolving a significant fraction of the cosmic submm
background into individual sources (\citet{smail97,hughes98}; see also 
review by  \citet{blain02} and 
references therein). A detailed characterization of these luminous
(sub)millimeter galaxies (SMGs) is only slowly accumulating
because of their faintness at all short wavelengths and the difficulty of 
counterpart identification. Photometric estimates of a median redshift of 
2.5-3 for the SMG population \citep[e.g.][]{carilli00} are 
consistent with the 
recent determination of a median redshift of z$\sim$2.2 for the 
$\lesssim$50\% of the
population accessible to optical spectroscopy \citep{chapman05}. The detection
of SMGs and the implied star formation rates and space densities 
immediately motivated speculation that they may trace the formation of
massive spheroids. Their central role in the assembly of massive 
galaxies is now becoming clearer through dynamical and through gas 
phase metallicity studies 
\citep{genzel03,neri03,tecza04,swinbank04,greve05}. Quantifying 
AGN in submillimeter galaxies is of immediate
relevance for determining their source of luminosity. It is of further
importance in the context of understanding the evolution of the
black hole mass to spheroid relation during one of the key phases of massive
galaxy formation, characterized by high star formation rates and gas content.

Rest frame mid-infrared spectroscopy can further our understanding of SMGs 
in two ways. First, it can determine redshifts for SMGs that have up to 
now eluded optical redshift measurements. Mid-infrared spectral features, 
in particular the narrow aromatic `PAH' features, if present, allow reasonably 
accurate redshift measurements even for targets that are extremely faint at 
optical wavelengths. Like the radio continuum commonly employed to locate
these galaxies \citep[e.g.][]{ivison02}, mid-infrared emission is closely 
linked to the bulk of the source luminosity that is emitted in the 
far-infrared, and in addition contains spectral features. The 
risk of erroneous redshift assignment due to misidentification is thus 
reduced. CO line emission shares this
property, however with existing instrumentation mid-infrared spectroscopy 
has the advantage of larger fractional bandwidth coverage than mm
spectroscopy. Second, low resolution mid-infrared 
spectroscopy can be used to constrain the energy sources and physical 
conditions of infrared-luminous galaxies, by decomposition into an 
AGN continuum and a
starburst component that is dominated by PAH emission features. This 
technique has been successfully applied
to the local infrared galaxy population during the ISO mission 
\citep[e.g.][]{genzel98, laurent00}. With the sensitivity of IRS on  the
Spitzer Space Telescope it is now possible to extend this method to high 
redshift infrared populations such as the SMGs. Identifying the mid-infrared 
AGN continuum by such spectral decomposition is possible even in the presence 
of significant obscuration, equivalent to tens of magnitudes in the rest frame
optical and more in the rest frame ultraviolet. With appropriate 
signal-to-noise, relatively faint continua can be identified that do not yet
strongly affect the broadband infrared colors that are another indicator of
AGN. An application of the infrared methods further constraining AGN 
properties is in combination 
with rest frame hard X-ray emission at levels signalling AGN activity, in 
particular for cases where the X-ray photon statistics is too limited to 
be able to fully constrain the X-ray obscuring column and thus the intrinsic
X-ray luminosity.

We are pursuing a program of Spitzer rest frame mid-infrared 
spectroscopy of a dozen bright and well studied SMGs. 
All have accurate interferometric positions that are the necessary 
prerequisite for such 
a study. In this letter, we present first results for two bright SMGs,
SMMJ02399-0136, the brightest source found in the 850$\mu$m SCUBA Cluster 
Lens Survey
\citep{smail97,smail02}, and MMJ154127+6616 detected near Abell 2125 in the
MAMBO 1.2mm survey \citep{bertoldi00}. We adopt $\Omega_m =0.3$,
$\Omega_\Lambda =0.7$ and
$H_0=70$ km\,s$^{-1}$\,Mpc$^{-1}$.

\section{Observations and Data Reduction}
Low resolution long slit spectra were obtained using Spitzer-IRS 
\citep{houck04} in the staring mode. SMMJ02399-0136 was observed in 
the LL1 19.5--38.0$\mu$m module for 30 cycles of 120sec ramp duration. 
The total on-source integration time was 2\,hours from addition of 
the independent 
spectra created by the telescope nod along the slit. MMJ154127+6616, without
an available optical redshift, was observed in the same way in the LL1 
module, and 
also for 15 cycles of 120sec ramp duration in the LL2 14.0--21.3$\mu$m 
module. We replaced deviant pixels in the individual differences of the two 
nod positions of 
the pipeline 11.0.2 basic calibrated data frames by values representative of
their spectral 
neighborhoods, and averaged the set of resulting 2-dimensional frames clipping
deviant values for the individual pixel fluxes.
After subtracting residual background emission, we used the SMART 
package \citep{higdon04} to extract calibrated 1-dimensional spectra for 
the positive and negative beams; these were averaged into the final spectra. 
Figs.~\ref{fig:j02spec} and \ref{fig:j15spec} show the extracted LL1 
spectra of SMMJ02399-0136 and MMJ154127+6616 as well as parts of the 
corresponding 2-dimensional frames. The shorter wavelength LL2 
spectrum of MMJ154127+6616 does not show significant continuum or feature 
emission at the target position, consistent with the tentative 
photometric fluxes
of 0.05mJy and 0.07mJy at 16$\mu$m and 22$\mu$m reported by 
\citet{charmandaris04}.

\section{Spectral Classification and Redshifts}

\begin{figure}
\epsscale{.80}
\plotone{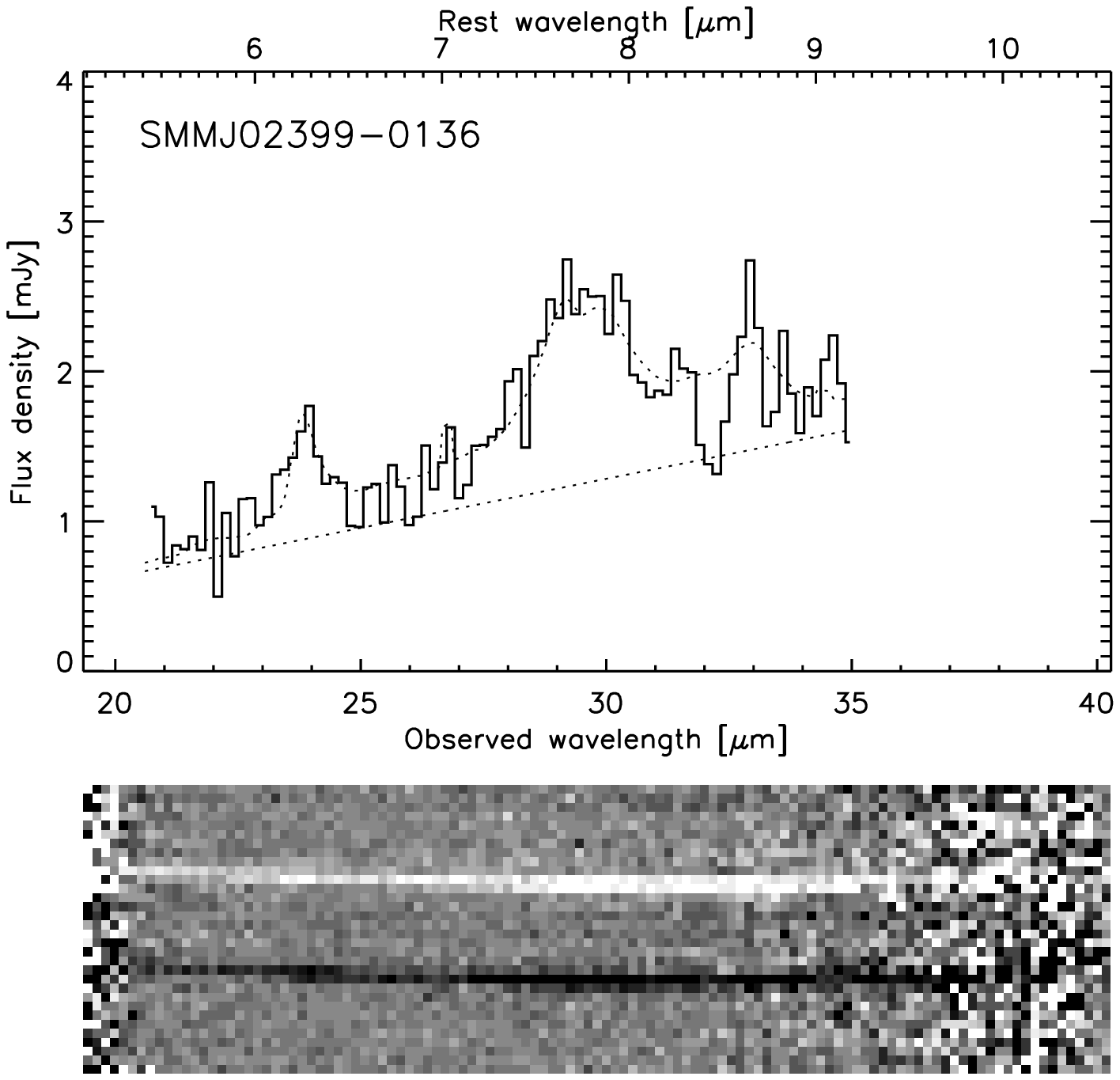}
\caption{Spitzer IRS low resolution spectrum of SMMJ02399-0136 (top) along with
the 2-dimensional long slit spectrum from which it was extracted (bottom). 
Note the increase of noise at long wavelengths. The
thin dotted lines indicate the fit of the spectrum by the sum of the scaled 
and redshifted ISO-SWS spectrum of the starburst M82, and a linearly sloped
and unabsorbed AGN continuum.}
\label{fig:j02spec}
\end{figure}

The spectrum of SMMJ02399-0136 shows well detected 6.2$\mu$m and 7.7$\mu$m 
aromatic 
`PAH' emission features superposed on a strong continuum. It is well fitted 
by the superposition of a scaled and redshifted starburst spectrum 
\citep[M82, ][]{sturm00} and a 
linearly rising, unabsorbed or little-absorbed continuum 
(Fig.~\ref{fig:j02spec}). This spectrum is 
similar to those of local universe infrared luminous
galaxies having significant contributions from both star formation and  
powerful AGN to their bolometric luminosities 
\citep[e.g. Mrk\,273,][]{genzel98}.
The global fit shown in Fig.~\ref{fig:j02spec} corresponds to a redshift of
2.83. The 6.2$\mu$m feature is intrinsically narrower and in a less complex 
part of the mid-infrared spectrum than the broader 7.7/8.6$\mu$m complex. 
Fitting just this feature and comparing to similar fits
to local universe ISO-SWS PAH spectra we obtain a redshift of 2.829 with a 
formal fit uncertainty of 0.013. Both 
values are in better than 1\% agreement with the accurate CO-based redshift 
of 2.8076 for SMMJ02399-0136 \citep{frayer98,genzel03}.

\begin{figure}
\epsscale{.80}
\plotone{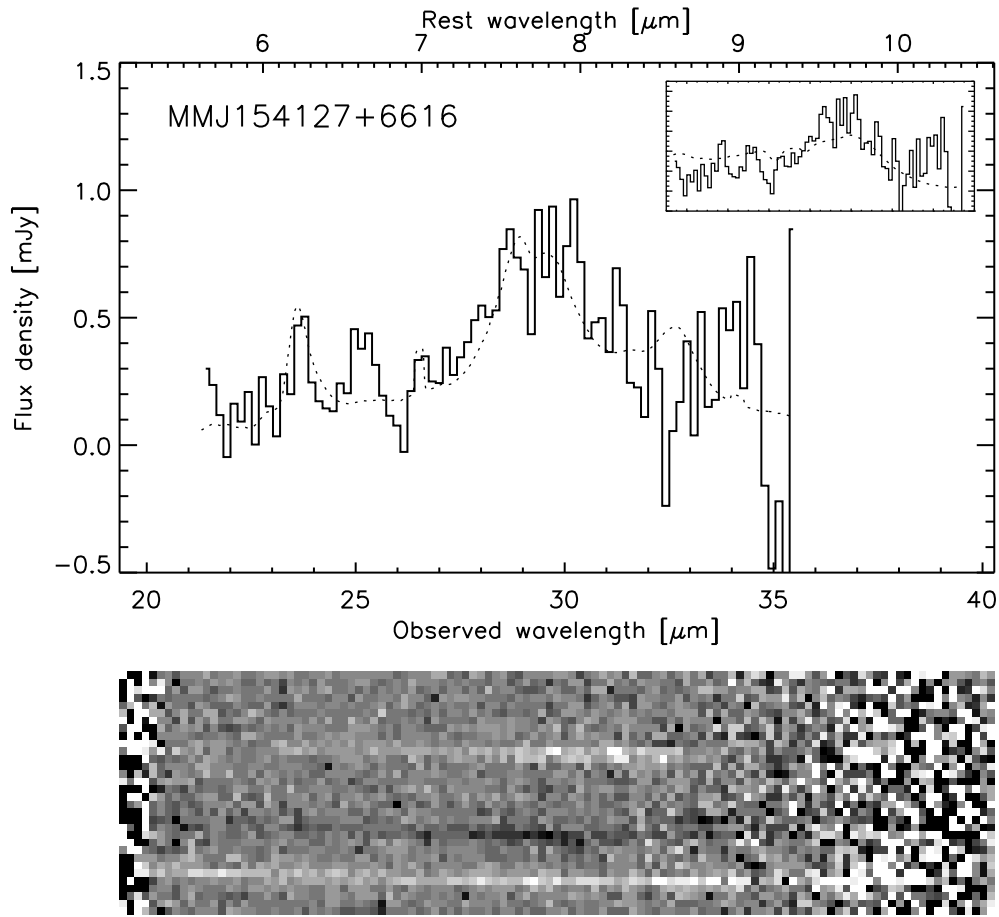}
\caption{Spitzer IRS low resolution spectrum of MMJ154127+6616. The thin 
dotted line indicates the fit by the scaled and redshifted spectrum of M82.
The small insert repeats the spectrum together with the lower quality fit 
by the obscured galaxy IRAS F00183-7111.  
The bright line at the bottom of the 2-dimensional spectrum is a serendipitous
source found in the slit.}
\label{fig:j15spec}
\end{figure}

The spectrum of the fainter MMJ154127+6616 is dominated by a broad peak 
near 30$\mu$m observed wavelength, with very weak emission at shorter and 
longer wavelengths. We identify this feature with the 
7.7$\mu$m PAH feature, based on the fit obtained to a redshifted starburst 
spectrum (Fig.~\ref{fig:j15spec}). We have also investigated fits
with the obscured continuum spectrum of IRAS F00183-7111 
\citep{tran01,spoon04}. Such spectra have a maximum near 
8$\mu$m rest wavelength, at the onset of the silicate absorption feature, 
and have to be considered as alternative fit to peaked mid-infrared spectra of
infrared galaxies.
While resulting in comparable redshifts, the fits with an F00183-like 
spectrum  do not reproduce the weakness or absence of short wavelength 
continuum emission (see insert of Fig.~\ref{fig:j15spec}). This is 
corroborated 
by a factor 1.5 increase in reduced $\chi^2$ of the 82 degree of freedom 
fit over this spectral range 
when changing from an M82 to an F00183 template. We suggest that the spectrum 
of MMJ154127+6616 is PAH dominated, consistent with the presence of a 
marginally significant maximum at the position of the 6.2$\mu$m feature.

To our knowledge, no redshift more accurate than the uncertain radio/submm 
estimates has been reported for MMJ154127+6616.
Using their SCUBA 850$\mu$m flux in addition to the MAMBO 1.2mm and radio data,
\citet[Object A2125-MM27 in their list]{eales03} estimate 
z=$2.45^{+0.67}_{-0.67}$ from the radio/submm relation
and z$<1$ from the technically difficult 1200$\mu$m to 850$\mu$m ratio. 
\citet{aretxaga03} estimate z=$2.8^{+1.7}_{-0.3}$ with slight variations for 
different adopted models.
From the fit in Fig.~\ref{fig:j15spec}, we derive z=2.80 with an 
uncertainty of $\Delta$z=0.1 estimated from the global fit. This uncertainty 
is likely
an upper limit given the indications for 6.2$\mu$m PAH emission. As for 
SMMJ02399-0136, this is somewhat larger than
 the median redshift for the part of the SMG population for which redshifts 
have been obtained in the optical 
\citep{chapman05}. These results demonstrate the power of 
IRS to derive redshifts not only for 24$\mu$m selected sources \citep{houck05},
but also for the those bright SMG sources where the 
redshifted mid-infrared features lie shortwards of the noisy $\lambda\gtrsim 
35\mu$m region of the IRS low resolution spectra.

For our adopted cosmology, a dust temperature of 45K and an emissivity 
index $\beta=1.5$ (consistent with the SMMJ02399-0136 studies of 
\citet{ivison98} and \citet{genzel03}) the intrinsic 8-1000$\mu$m luminosity is
$L_{IR}\sim1.2\times10^{13} L_\odot$ for SMMJ02399-0136, after correcting 
for the lensing amplification of 2.45. We estimate 
$L_{IR}\sim1.9\times10^{13} L_\odot$ for MMJ154127+6616 scaling with the 
intrinsic 850$\mu$m fluxes.

\section{ULIRG-like rest frame mid- to far-infrared SEDs}

\begin{figure}
\epsscale{.80}
\plotone{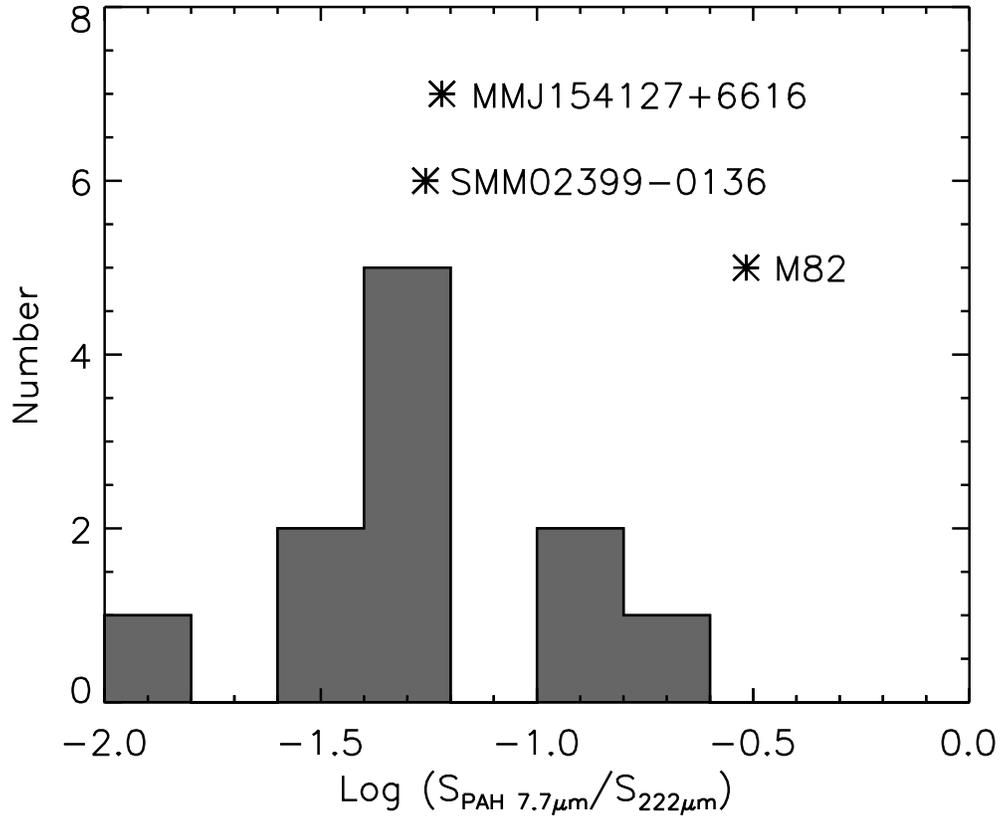}
\caption{Histogram showing the ratio of PAH 7.7$\mu$m peak flux density and
rest frame 222$\mu$m continuum flux density for eleven local ULIRGs. In this
measure of the mid- to far-infrared SED, the two SMGs are very similar to 
the local ULIRG population. However, both show a lower value than the low 
luminosity starburst M82. The M82 point is based on the PAH 
data of \citet{foerster03} and the far-infrared continuum of 
\citet{colbert99}, obtained in large and similar apertures.}
\label{fig:pahfir}
\end{figure}

The rest frame mid- to far-infrared SEDs provide clues 
about properties, energy sources, and radiation fields of SMGs but are 
presently insufficiently 
constrained by direct observations. Constraints on their
far-infrared peaks are currently indirect using the observed submm and 
radio fluxes and the assumption of the radio/far-infrared correlation 
for star forming galaxies \citep{chapman03, chapman05}. In view of the 
variations observed in local galaxy SEDs, we have added another 
constraint by comparing the ratio of
PAH features and SCUBA 850$\mu$m continua (rest frame $\sim222\mu$m for 
our two objects). We adopt SCUBA fluxes of 23mJy for SMMJ02399-0136 
\citep{smail02} and 14.6mJy for MMJ154127+6616 \citep[A2125-MM27 in][]{eales03}.
In Fig.~\ref{fig:pahfir} the ratio of peak flux density of the 7.7$\mu$m PAH 
feature after continuum subtraction to the continuum flux density at 
rest frame 222$\mu$m for our two SMGs is compared to 
the same measure for 11 local ultraluminous infrared galaxies (ULIRGs) with 
PAH emission, where the PAH data 
have been taken from ISOPHOT-S observations \citep[e.g.][]{rigopoulou99} and 
the continua from slight
extrapolations of the far-infrared photometry of \citet{klaas01}, which 
reaches 
out to 200$\mu$m observed wavelength. The PAH to far-infrared ratios
of the two SMGs are fully consistent with that of the local ULIRG population.
These SED properties are in agreement with the finding from spatially 
resolved mm interferometry that SMGs are scaled up versions of the compact
star formation events in local ULIRGs \citep{tacconi05}.
 
\section{AGN content}

With both strong AGN continuum and PAH features, at a feature-to-continuum 
ratio $\sim$1, SMMJ02399-0136 is at the transition between predominantly 
starburst powered and predominantly AGN powered according to the mid-infrared 
diagnostic of \citet{genzel98}. AGN signatures are also seen in the optical
spectrum \citep{ivison98}. The combination of mid-infrared spectroscopy 
and the Chandra X-ray data of \cite{bautz00} can constrain the properties of 
this AGN. \citet{bautz00} clearly detect relatively hard X-ray emission 
from SMMJ02399-0136, with an observed luminosity in the rest frame 2-10keV band
of $0.18\times 10^{44}$erg s$^{-1}$ (corrected to our adopted cosmological 
parameters). Because of limited photon statistics, their data cannot
discriminate between seeing the direct AGN emission, although absorbed by a 
significant column of
$N_H\sim 10^{24}$cm$^{-2}$ (with the intrinsic emission $\sim$15 times 
brighter), and a fully Compton thick AGN seen in reflection, with the 
intrinsic emission brighter by the inverse of their adopted reflection 
efficiency of 0.022. 
For two related reasons, the Compton thick case appears more 
consistent with our infrared 
observations. First, the ratio of rest frame hard X-rays
and mid-infrared AGN continuum 
$log(L_{2-10keV}/\nu L_\nu(6\mu m))$ is -0.54 for the 
reflected case. This ratio is at the center of the equivalent 
distribution of mid-infrared to 
intrinsic hard X-ray ratios for local AGN, studied by \citet{lutz04}, while
the lower ratio for the direct emission case would be a the lower end of this 
distribution. Second, 
correcting from intrinsic hard X-rays to AGN bolometric luminosity 
assuming $L_{2-10keV}/L_{Bol}\sim 0.09$ for AGN \citep{elvis94}, the reflected
case with L$_{Bol,AGN}\approx 2\times 10^{12}L_\odot$ is closer to the roughly 
similar contributions of AGN and star formation that are suggested by the 
infrared spectroscopy. In summary, we conclude that SMMJ02399-0136 is powered 
by roughly equal contributions of star formation and a Compton-thick AGN.

No strong AGN continuum is present in MMJ154127+6616 (Fig.~\ref{fig:j15spec}),
in particular below $\sim$6.5um rest wavelength where it would be most easily 
detectable in the presence of PAHs. This is further supported by the 
weakness of the tentative 16 and
22$\mu$m photometric detections of \citet{charmandaris04}, at levels below 
0.1mJy. 
As is the case for local ULIRGs, it is appropriate to caution
that our conclusion of dominance of star formation does not exclude in 
any way the presence of an AGN that is a very minor contributor to the 
bolometric luminosity. A significant number of minor AGN in SMGs is suggested
by the deepest Chandra observations \citep[]{alexander03,alexander05}.

\section{Discussion} 

We have presented Spitzer mid-infrared spectroscopy of two of the 
brightest known submillimeter galaxies, SMMJ02399-0136 and MMJ154127+6616. 
Our unambiguous detections of PAH spectral features and mid-infrared
continua allows us to constrain the energy sources in these objects, and 
to determine the previously unknown redshift for one of them (MMJ154127+6616).
We find that the luminosity of the first galaxy is generated by 
approximately equal contributions from star-formation and an AGN.
The second galaxy is dominated by star formation.

The existence of star formation dominated systems at infrared luminosities in 
excess of 10$^{13}$L$_\odot$ is unique to the high redshift universe.
In our previous ISO studies of ULIRGs in the local universe, using the 
same mid-infrared methods, we have found star formation dominated 
systems only up to a luminosity of 10$^{12.65}$L$_\odot$ 
\citep{rigopoulou99,tran01}. Similar 
conclusions have been reached from optical spectroscopy of local ULIRGs
\citep{veilleux99}. The existence of higher luminosity starbursts in SMGs 
may be related to their higher gas fractions \citep{greve05,tacconi05}.
Star formation proceeding at these extreme rates
in high redshift objects ($\gtrsim$1000M$_\odot$/yr for 
$L_{IR}>10^{13}$L$_\odot$) naturally fits into the evolving
understanding of the formation of massive galaxies at high redshift, however. 
Such extreme events can trace the formation of the 
10$^{11}$M$_\odot$ galaxies already fully assembled at redshifts z=1.6 to 1.9
\citep{cimatti04}. Rapid star-formation may also meet the constraint on rapid 
formation of massive ellipticals inferred from measurements of the 
$\alpha$/Fe element abundance ratios \citep{thomas05}.

Spectroscopy with IRS can play a central role in elucidating the relationships
amongst various infrared selected high redshift galaxy 
populations. \citet{houck05} have recently obtained IRS spectra of 24$\mu$m
bright ($>$0.75mJy) but optically faint (R$>$24) galaxies selected from a 
large area Spitzer survey. Submillimeter galaxies at z$\sim$2 
with an Arp220-like SED but still more luminous than a typical SMG from the 
current SCUBA/MAMBO surveys could meet 
the basic brightness and obscuration constraint of this sample. The 
preponderance of heavily obscured continua in the \citet{houck05} spectra,
however, 
unlike the PAH and PAH plus continuum spectra of our two SMGs, argues for a 
small overlap between the two populations, and that the population of 
R-band faint but 24$\mu$m bright sources may be dominated by obscured AGN.

\acknowledgments

We thank Natascha F\"orster Schreiber for access to ISOCAM spectra of M82 
and the referee for helpful comments.

%Facilities: \facility{Spitzer(IRS)}

%up to 8 authors, then first et al.

\clearpage

\clearpage

\clearpage

\clearpage

\end{document}